\documentclass[aps,prl,twocolumn,superscriptaddress,showpacs,floatfix]{revtex4}
\usepackage{graphicx,amsmath,amssymb,bm}

\newcommand{\be}{\begin{equation}}
\newcommand{\ee}{\end{equation}}
\newcommand{\vlk}{V_{\text{low}\,k}}

\newcommand{\la}{\Lambda}

\begin{document}

\title{Low-momentum interaction in few-nucleon systems}
\author{Andreas Nogga}
\email[E-mail:~]{nogga@phys.washington.edu}
\affiliation{Institute for Nuclear Theory, Box 351550, University
of Washington, Seattle, WA 98195}
\author{Scott K. Bogner}
\email[E-mail:~]{bogner@phys.washington.edu}
\affiliation{Institute for Nuclear Theory, Box 351550, University
of Washington, Seattle, WA 98195}
\author{Achim Schwenk}
\email[E-mail:~]{aschwenk@mps.ohio-state.edu}
\affiliation{Department of Physics, The Ohio State University,
Columbus, OH 43210}

\date{\today}

\begin{abstract}
The effective low-momentum interaction $\vlk$ is applied to 
three- and four-nucleon systems. We investigate the
$^3$H, $^3$He and $^4$He binding energies for a wide range of 
the momentum cutoffs. By construction, all low-energy two-body
observables are cutoff-independent, and therefore, any cutoff dependence is
due to missing three-body or higher-body forces.
We argue that for reasonable cutoffs $\vlk$ is similar to high-order
interactions derived from chiral effective field theory. This 
motivates augmenting $\vlk$ by corresponding three-nucleon forces. 
The set of low-momentum two- and three-nucleon forces can be used 
in calculations of nuclear structure and reactions.
\end{abstract}

\maketitle

Microscopic nuclear many-body calculations are complicated by the   
short-distance repulsion in nuclear forces, which leads to strong 
high-momentum components in nuclear wave functions. Usually, 
one solves this problem by introducing an effective interaction, 
the Brueckner $G$ matrix, which resums in-medium particle-particle 
scattering. The $G$ matrix is a soft interaction, which is both
energy- and nucleus-dependent and typically requires approximations
in practice.  Moreover, the resummation of
particle-particle contributions makes it extremely complicated
to treat particle-particle and particle-hole  
correlations on an equal footing.

An alternative strategy to construct a soft interaction by integrating out
the high-momentum components in free space
has been formulated in~\cite{bogner03a}. Using a renormalization 
group (RG) approach, phenomenological two-body potential models can 
be evolved to an effective low-momentum interaction, called $\vlk$, 
which is energy-independent, hermitian and preserves the on-shell 
$T$ matrix below a cutoff $\la$ in momentum space as well as 
the deuteron binding energy. For $\la \lesssim 2 \, \text{fm}^{-1}$, 
the matrix elements of $\vlk$ are practically independent of the 
potential model it is derived from and thus unifies all nuclear forces 
used in microscopic nuclear structure calculations~\cite{bogner03b}. 
By construction, $\vlk$ is much softer than the modern potential models, 
and thus can be used directly for microscopic nuclear calculations in 
different mass regions~\cite{bogner02,coraggio03} or for different 
densities~\cite{schwenk03,schwenk04}. This is 
clearly important to theoretically 
extrapolate to the nuclear drip lines without ambiguities due to 
unknown interactions.

Over the last few years, there has also been an immense 
progress in our understanding of nuclear interactions from chiral effective 
field theory (EFT). This approach qualitatively explains the 
hierarchy of two-nucleon (2N), three-nucleon (3N) and higher-body 
forces~\cite{weinberg90}, which is observed using phenomenological 
models. On a quantitative level, it was shown that the resulting 
2N and consistent higher-body interactions lead to a quite good
description of 2N as well as 3N 
observables~\cite{ordonez96,epelbaum00,entem03a,kolck94,epelbaum02c}. 
In the pionfull EFT approach, the Lippmann-Schwinger equation is 
regularized by imposing a cutoff $\la \approx 2.5 - 3.0 \, 
\text{fm}^{-1}$. Thus, the chiral potentials are also low-momentum 
interactions, and with the universal property of $\vlk$, this suggests 
that $\vlk$ effectively parameterizes higher-order chiral 2N interactions.
While EFT offers the only systematic approach to consistent 2N and 
higher-body forces, $\vlk$ can be evolved to arbitrary cutoffs with 
cutoff-independent 2N observables.

Since $\vlk$ is constructed within the 2N system, one neglects many-body 
forces due to degrees of freedom missing in the effective theory
(contributions from the $\Delta$) as well as due the truncation to low 
momenta (contributions from high-momentum nucleons). In any effective
theory, these effects are inseparable.  In this Letter, we use cutoff
dependence as a tool to assess the effects of many-body forces.
Motivated by the similarities between $\vlk$ and chiral low-momentum
interactions, we combine $\vlk$ with the leading chiral 3N force 
to absorb the cutoff dependence in $A \leqslant 4$ binding energies. Finally, 
we examine the expectation values of the various force components
to check that the hierarchy of nuclear two- and three-body forces is 
maintained.

We first calculate 3N and 4N binding energies by solving the
Faddeev-Yakubovsky equations with only the two-body $\vlk$. 
We include electromagnetic and isospin-breaking effects and
vary the cutoff over a wide range. Our results are numerically 
stable for the studied cutoff values, which requires a careful 
treatment of the necessary interpolations in the vicinity of 
the sharp cutoff. We also checked the convergence with respect 
to the included partial waves. We estimate an accuracy 
of $2 \, \text{keV}$ for the $^3$H and $^3$He and $50 \, \text{keV}$ 
for the $^4$He calculations. More details about the numerical method 
can be found in~\cite{nogga02b}.

\begin{figure}[btp]
\begin{center}
\includegraphics[width=7.5cm,clip=]{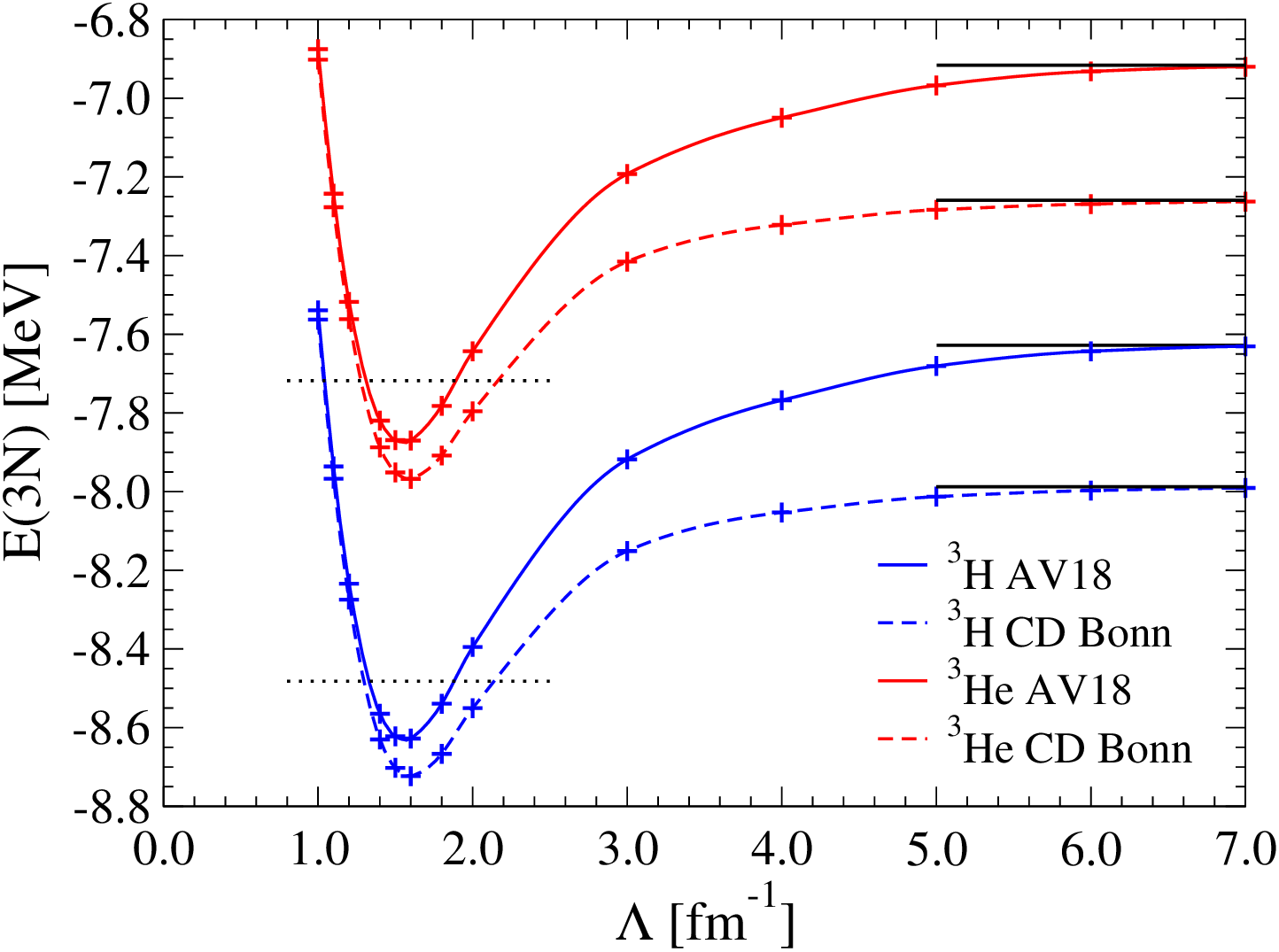}
\end{center}
\vspace{-0.5cm}
\begin{center}
\includegraphics[width=7.5cm,clip=]{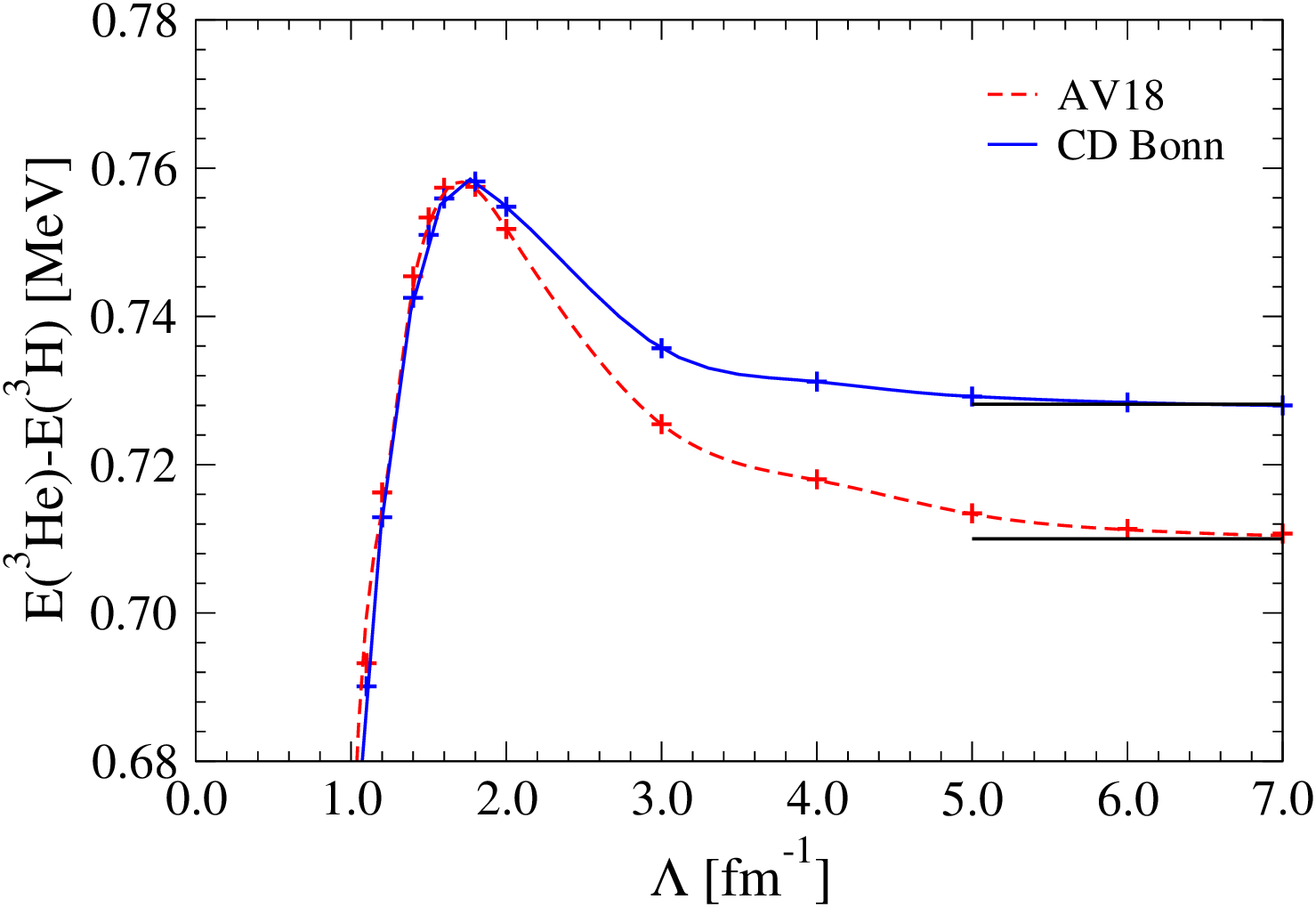}
\end{center}
\vspace{-0.6cm}
\caption{\label{fig:3n} (Color online) Cutoff dependence of the 3N binding 
energies and the binding energy difference of $^3$H and 
$^3$He. Results are shown for the Argonne $v_{18}$ and 
the CD-Bonn~2000 potential. The horizontal solid lines represent 
results for the bare two-body interactions and the dotted lines
denote the experimental binding energies.}
\end{figure}

In Fig.~\ref{fig:3n}, we give results for binding energies of the
3N system. We show results for the
$\vlk$ derived from the CD-Bonn~2000~\cite{machleidt01a} and 
Argonne $v_{18}$~\cite{wiringa95} interactions. The cutoff
dependence is due to missing three-body forces. For large cutoffs,
we reproduce the known binding energies obtained with the bare 
interactions only. For intermediate cutoffs, we find a stronger 
binding with $\vlk$.  This could be expected, because softer 
interactions generally lead to stronger binding. It is also 
consistent with the correlation between the triton binding energy 
and the deuteron D-state probability observed for phenomenological 
potentials~\cite{afnan75}. For $\vlk$ the D-state probability decreases
monotonically with a decreasing cutoff. Therefore, this 
correlation evidently breaks down for cutoffs below 
$\la \approx 1.6 \, \text{fm}^{-1}$. The binding then
decreases, as attractive 
parts of the bare interactions are integrated out.

For cutoffs $\la \lesssim 2 m_\pi$, truly model-independent 
results are obtained and the binding energy curves for the 
CD-Bonn~2000 and Argonne $v_{18}$ $\vlk$ interactions collapse. 
In Fig.~\ref{fig:3n}, we also show the cutoff dependence of the 
difference in $^3$He and triton binding energies, which is due 
to electromagnetic and isospin-breaking contributions. The difference 
varies by $60 \, \text{keV}$ and correlates with the binding energy, 
since the latter is related to the charge radius~\cite{friar87}.
For special choices of the cutoff, both 
experimental binding energies can be reproduced simultaneously 
without a 3N interaction. We emphasize that 3N forces will contribute 
to other observables. Nevertheless, it may be interesting to study 
many-body systems using these particular values for the momentum cutoff, 
since a simple zero-range 3N force vanishes in these cases.

Our results indicate that 3N forces due to the truncation to low 
momenta are of the same order as adjusted 3N forces due to missing 
excitations of nucleons, although these  effects cannot be separated. 
The bare 3N forces provide about $0.7 - 1 \, \text{MeV}$ of binding 
in conventional models, whereas the binding energies given by $\vlk$
change by $1 \, \text{MeV}$ over the large cutoff range.
In this sense the truncation to low momenta does not induce strong
three-body forces. We note that this is in contrast to the 
interpretation given in~\cite{fujii04}. There, the size of 3N forces 
was assessed by comparing the $\vlk$ binding energies to the results 
of the bare 2N potential model. This neglects the uncertainty in the 
binding energy predictions of traditional 2N forces and misses that, 
in effective theory approaches, the effects of the truncation 
to small cutoffs are inseparable from those of missing degrees 
of freedom like the $\Delta$. Because these two contributions 
to higher-body forces cannot be disentangled at low energies,  
we will absorb both by augmenting $\vlk$ with a chiral 3N force below.

\begin{figure}[btp]
\begin{center}
\includegraphics[width=7.5cm,clip=]{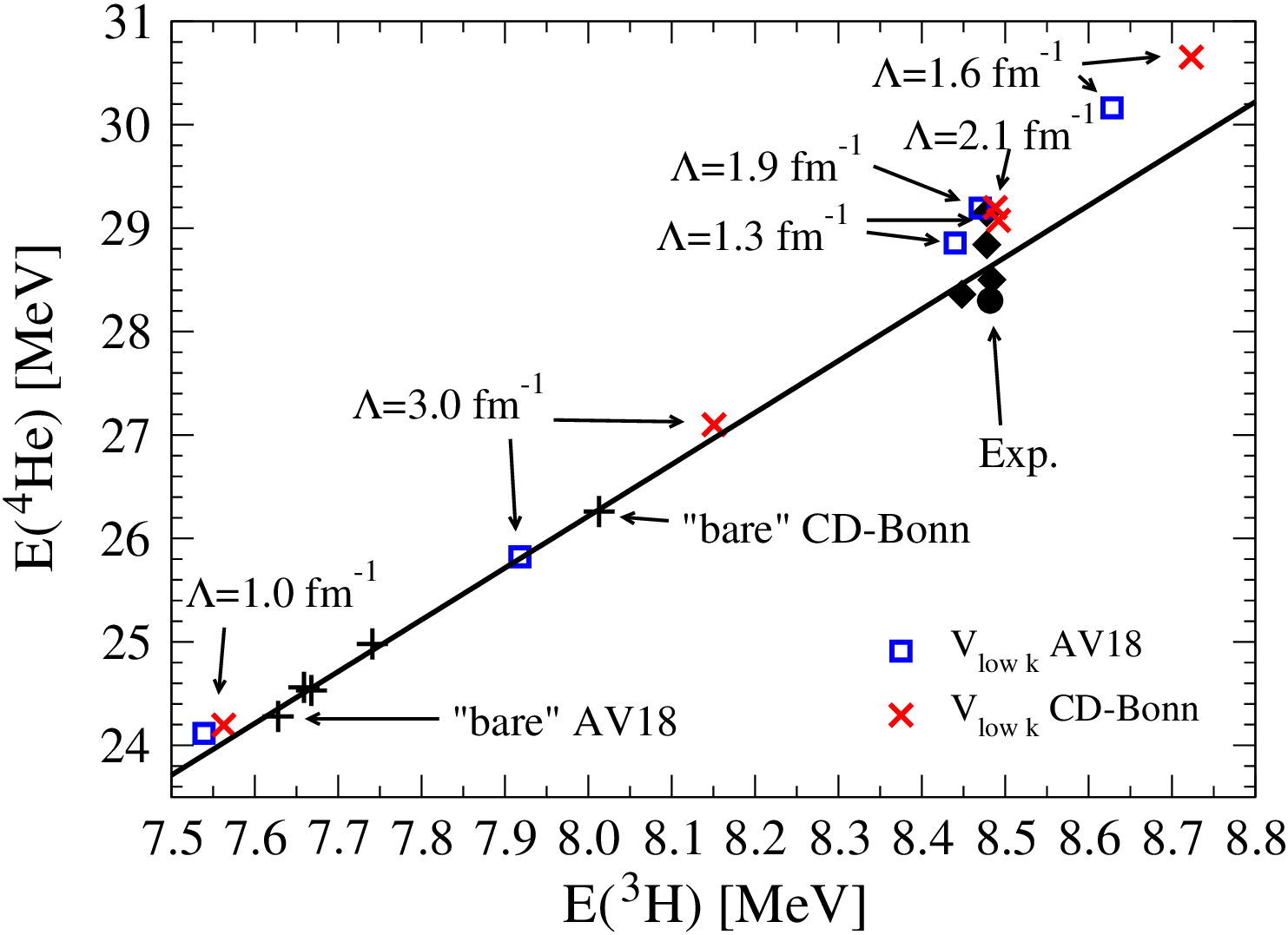}
\end{center}
\vspace{-0.5cm}
\caption{\label{fig:tjon} (Color online) 
Correlation of the $^3$H and $^4$He binding energies. Results are shown for 
several modern potential models alone (plusses) and with adjusted 3N forces 
(diamonds)~\cite{nogga00}. The $\vlk$ results are for the Argonne $v_{18}$ 
(squares) and the CD-Bonn~2000 potential (crosses). The solid line is a 
linear fit to the 2N force model results only.}
\end{figure}

For further insight, we have calculated the $\alpha$-particle binding
energy. To obtain an overview, calculations are performed for the 
smallest cutoff considered $\la = 1.0 \, \text{fm}^{-1}$, 
in the maximum of the triton binding energy at 
$\la = 1.6 \, \text{fm}^{-1}$, for two cutoffs 
which lead to 3N binding energies close to the experimental one,
$\la = 1.3 \, \text{fm}^{-1}$ and $\la= 1.9 \, \text{fm}^{-1}$ 
(Argonne $v_{18}$) or $\la=2.1 \, \text{fm}^{-1}$ (CD-Bonn~2000), 
and for a cutoff in the tail at $\la = 3.0 \, \text{fm}^{-1}$.   
The focus of our studies is whether the cutoff dependence of
$\vlk$ can be related to correlations observed  when traditional two-body
interactions are used. From~\cite{tjon75,nogga00,nogga02b},
it is well-known that there is an almost linear relation
between 3N and 4N binding energies, known as the Tjon-line. This
correlation holds with very good accuracy for all modern interactions, 
but is slightly broken by the action of 3N forces. As can be seen in 
Fig.~\ref{fig:tjon}, the various $\vlk$ results do not differ 
significantly more from the phenonemological Tjon-line than calculations 
with adjusted 3N forces. We see that as a further indication that 
3N and 4N contributions are not unexpectedly large due to the 
low-momentum truncation, at least for the triton and $\alpha$-particle.
Already at $\la = 3.0 \, \text{fm}^{-1}$ the $\vlk$ prediction is 
almost exactly on the Tjon-line given by the phenomenological 
models. 

\begin{table}[btp]
\begin{center}
\begin{tabular}{l|rr}
$\la \, [\text{fm}^{-1}]$ & $c_D$ & $c_E$ \\ \hline 
$1.0$ & $3.621$ & $5.724$ \\
$1.3$ & $11.889$ & $2.265$ \\
$1.6$ & $2.080$ & $0.230$ \\
$1.9$ & $-1.225$ & $-0.405$ \\
$2.5(a)$ & $-0.560$ & $-0.707$ \\
$2.5(b)$ & $-3.794$ & $-1.085$ \\
$3.0(*)$ & $-7.500$ & $-2.151$ 
\end{tabular}
\end{center}
\caption{\label{tab:fit} Fit results for $c_D$ and $c_E$ for various 
cutoffs of the $\vlk$ derived from the Argonne $v_{18}$ potential (for 
$(*)$ see text). The strength of the $2 \pi$ exchange 
part is determined by $c_1 = -0.76 \, \text{GeV}^{-1}$, $c_3 = -4.78 
\, \text{GeV}^{-1}$ and $c_4 = 3.96 \, \text{GeV}^{-1}$~\cite{rentmeester03}.}
\end{table}

\begin{table*}[btp]
\begin{center}
\begin{tabular}{l|rrrrr|rrrrr}
& \multicolumn{5}{c|}{$^3$H} & \multicolumn{5}{c}{$^4$He} \\
\multicolumn{1}{c|}{$\la \, [\text{fm}^{-1}]$} &
\multicolumn{1}{c}{$T$} & \multicolumn{1}{c}{$\vlk$} &
\multicolumn{1}{c}{$c$-terms} & \multicolumn{1}{c}{$D$-term} &
\multicolumn{1}{c|}{$E$-term} & \multicolumn{1}{c}{$T$} &
\multicolumn{1}{c}{$\vlk$} & \multicolumn{1}{c}{$c$-terms} &
\multicolumn{1}{c}{$D$-term} & \multicolumn{1}{c}{$E$-term} \\ \hline
$1.0$ & $21.06$ & $-28.62$ & $0.02$ & $0.11$ & $-1.06$ & 
$38.11$ & $-62.18$ & $0.10$ & $0.54$ & $-4.87$ \\
$1.3$ & $25.71$ & $-34.14$ & $0.01$ & $1.39$ & $-1.46$ & 
$50.14$ & $-78.86$ & $0.19$ & $8.08$ & $-7.83$ \\
$1.6$ & $28.45$ & $-37.04$ & $-0.11$ & $0.55$ & $-0.32$ & 
$57.01$ & $-86.82$ & $-0.14$ & $3.61$ & $-1.94$ \\
$1.9$ & $30.25$ & $-38.66$ & $-0.48$ & $-0.50$ & $0.90$ &
$60.84$ & $-89.50$ & $-1.83$ & $-3.48$ & $5.68$ \\ 
$2.5(a)$ & $33.30$ & $-40.94$ & $-2.22$ & $-0.11$ & $1.49$ &
$67.56$ & $-90.97$ & $-11.06$ & $-0.41$ & $6.62$ \\
$2.5(b)$ & $33.51$ & $-41.29$ & $-2.26$ & $-1.42$ & $2.97$ &  
$68.03$ & $-92.86$ & $-11.22$ & $-8.67$ & $16.45$ \\
$3.0(*)$ & $36.98$ & $-43.91$ & $-4.49$ & $-0.73$ & $3.67$ & 
$78.77$ & $-99.03$ & $-22.82$ & $-2.63$ & $16.95$
\end{tabular}
\end{center}
\caption{\label{tab:expt} Expectation values of the kinetic energy ($T$), 
2N interaction ($\vlk$), $2 \pi$ exchange part of the 3N force ($c$-terms)
and $D$- and $E$-term for $^3$H and $^4$He. All energies are in MeV.}
\end{table*}

\begin{figure}[btp]
\begin{center}
\includegraphics[width=7.5cm,clip=]{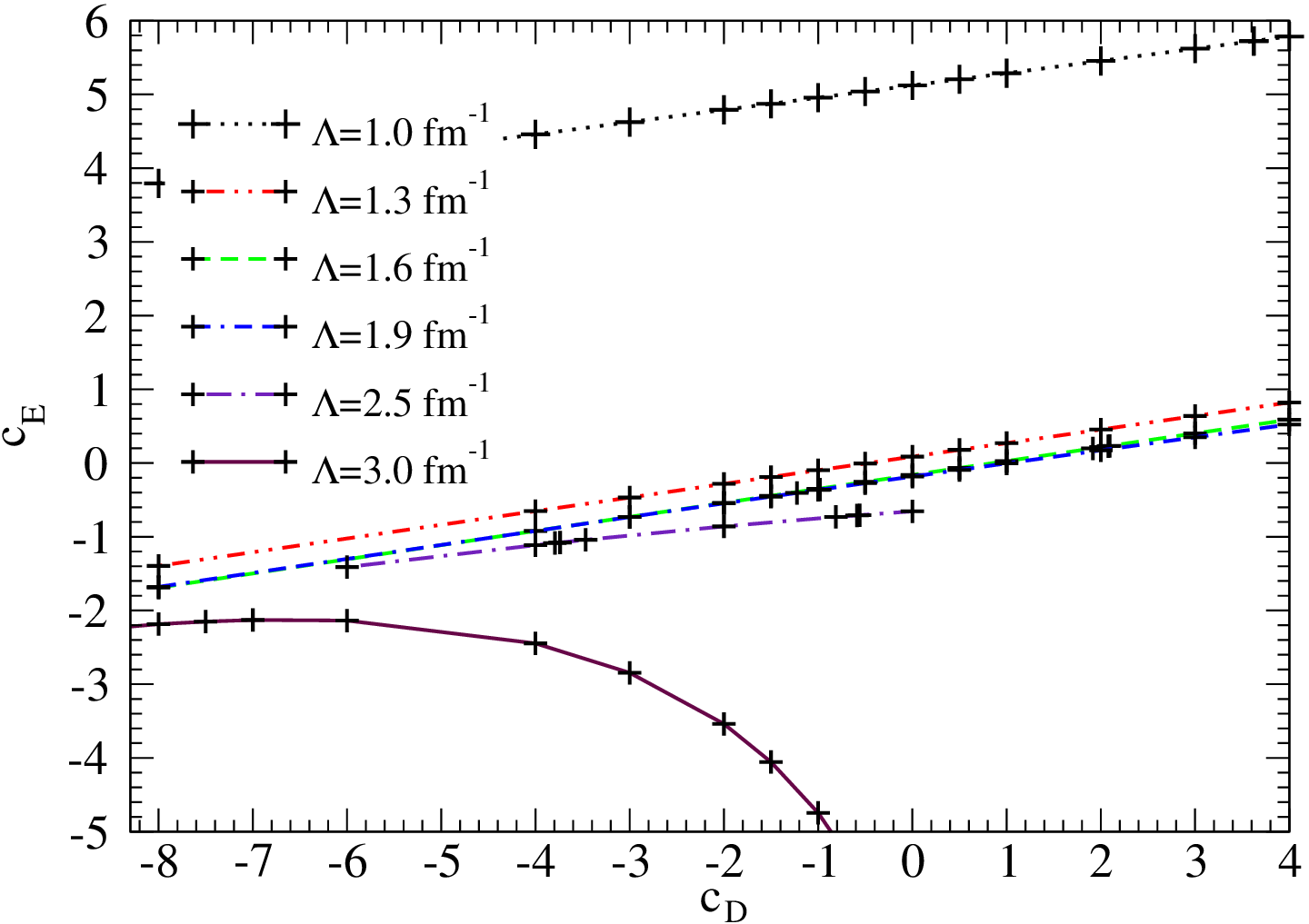}
\end{center}
\vspace{-0.5cm}
\caption{\label{fig:fit3h} (Color online) Relation between $c_D$ and $c_E$ 
obtained by requiring that $\vlk$ augmented by the 3N force predicts the 
$^3$H binding energy correctly.}
\end{figure}

As also seen in Fig.~\ref{fig:tjon}, even if a cutoff is chosen that leads 
to a good description of the 3N binding energies, the 4N binding energy
deviates from experiment. Clearly, 3N or higher-body forces must act for 
these values of the cutoff. In the following, we construct a low-momentum 
3N interaction by fitting the leading chiral 3N force to $\vlk$. For
simplicity we restrict ourselves to the $\vlk$ derived from the Argonne
$v_{18}$ potential. The chiral 3N force to leading order contains a 
long-range $2 \pi$ exchange part, an intermediate range one $\pi$ 
exchange ($D$-term) and a zero-range contact interaction ($E$-term), 
see~\cite{kolck94,epelbaum02c}. For the operator form and the definition 
of the strength constants, we refer the reader to Eqs.~(2) and (10) 
in~\cite{epelbaum02c}. The interaction is regularized by exponential 
cutoff functions of the form $\exp(-\left(p/\Lambda)^8\right)$
with the cutoff taken from $\vlk$. The very high exponent guarantees
a very sharp drop to zero at $p=\Lambda$. The $2 \pi$ exchange 
part is determined by strength constants $c_i$, which we take 
from~\cite{rentmeester03}, where they were obtained by a fit to 
NN data.
 The dimensionless strength 
constants $c_D$ and $c_E$ were obtained from a fit to the $^3$H and 
$^4$He binding energies. First, a relation between $c_D$ and $c_E$ was 
established by requiring that the $^3$He binding energy of $-8.482 \,
\text{MeV}$ is described accurately. The resulting dependence for various 
cutoffs is shown in Fig.~\ref{fig:fit3h}. For small cutoffs we obtain a 
linear relationship, which suggests that the 
$D$- and $E$-terms are perturbative 
in this region. We have checked explicitly and also for the $c$-terms 
that these are perturbative for $\la \lesssim 2 \, \text{fm}^{-1}$. 
This could be useful for applications, where it is practically
impossible to include the 3N force into the dynamical equations, but a 
perturbative treatment is feasible. 

\begin{figure}[btp]
\begin{center}
\includegraphics[width=7.5cm,clip=]{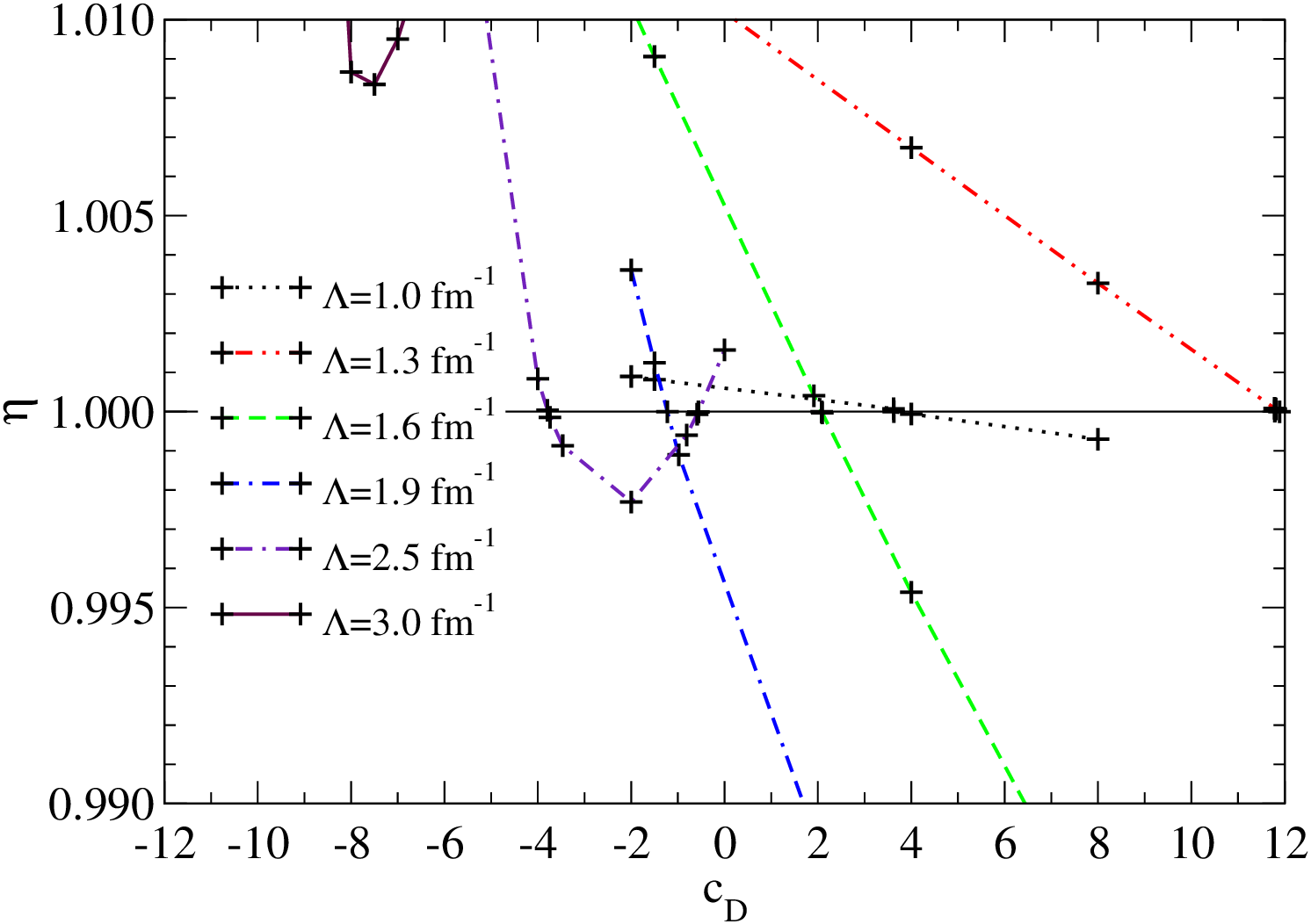}
\end{center}
\vspace{-0.5cm}
\caption{\label{fig:fit4he} (Color online) Dependence of the eigenvalue 
$\eta$ of the Yakubovksy equation on $c_D$ for various cutoffs. A deviation 
of $\eta-1 = 0.01$ corresponds to a deviation of approximately $600 \,
\text{keV}$ from the experimental value.}
\end{figure}

In Fig.~\ref{fig:fit4he}, we show the eigenvalue $\eta$ of the Yakubovsky 
equation for $^4$He versus $c_D$. In all cases $c_E$ was chosen 
according to Fig.~\ref{fig:fit3h}. The binding energy of 
$^4$He agrees with the experimental one of $-28.3 \, \text{MeV}$ for 
$\eta=1$. In the considered range for $c_D$, we find a unique solution 
for the cutoff choices up to $\la = 1.9 \, \text{fm}^{-1}$. For 
$\la = 2.5 \, \text{fm}^{-1}$, the relation of $\eta$ and $c_D$ is 
strongly non-linear and we find two solutions. We observed a very similar 
behavior, when the N3LO chiral interaction of~\cite{entem03a} was 
augmented by the same 3N force. For $\la = 3.0 \, \text{fm}^{-1}$, 
we cannot describe the $^3$H and $^4$He binding energies simultaneously. 
For this cutoff, we choose $c_D = 7.5$, for which $\eta$ is minimal and 
the binding energy is best described. The resulting $c_D$/$c_E$ pairs are 
compiled in Table~\ref{tab:fit}, where the $(*)$ indicates that the
$^4$He binding energy is reproduced only approximately as $-28.8 \, 
\text{MeV}$ for $\la = 3.0 \, \text{fm}^{-1}$.

A very important task is to estimate the size of 3N forces in a 
systematic way. We decided to calculate the expectation values of 
the 2N and the different parts of the 3N interactions and compare 
their magnitude. The results are summarized in Table~\ref{tab:expt}.
As a worst case scenario, we compare the maximum of the individual 
3N force terms to the 2N interaction for $^4$He. As expected from 
Fig.~\ref{fig:fit3h}, for $\la \lesssim 2 \, \text{fm}^{-1}$, all 3N 
parts are perturbative. For these cutoffs, we obtain contributions of 
$4 - 10 \%$, which is comparable to 3N forces for phenomenological 
models~\cite{nogga02b,carlson98}. For larger cutoffs, the $2 \pi$ 
exchange contribution ($c$-terms) grows rapidly, which is canceled 
by the $E$-term. We take this as an indication that, in this range, 
our ansatz for the 3N force is not reliable.

In summary, we have thoroughly assessed the size of 3N forces in the 
$\vlk$ approach. Based on the $\vlk$ results for the $^3$H and $^4$He binding 
energies, we found that the dependence on the cutoff is not unnaturally
large for $\Lambda \geqslant 1.0 \, \text{fm}^{-1}$.  This suggests that 
higher-body interactions are small.
We emphasize that the large cutoff range, for which $\vlk$ is available, will 
enable similar studies for other low-energy observables, e.g., all binding 
and excitation energies, and that this is a powerful tool to isolate missing 
parts in effective interactions. Furthermore, 
we have extended $\vlk$ by the 
leading chiral 3N force and fitted the two unknown parameters to the 
$^3$H and $^4$He binding energies. We assessed the strength of the 3N force 
by calculating expectation values of its individual parts. By requiring 
that not only the sum, but also the individual parts are of natural size, 
we found that our ansatz for the 3N force is reliable for cutoffs $\la 
\lesssim 2 \, \text{fm}^{-1}$. It turned out that the 3N force contribution 
can be treated perturbatively for this range of cutoffs. This completes
a soft nuclear interaction model, which will be important for many-body 
calculations. Applications to symmetric nuclear matter are in
preparation. 

We are grateful to Dick Furnstahl for useful discussions. This work
was supported by the US-DOE under Grants No. DE-FC02-01ER41187,
DE-FG02-00ER41132 and the NSF under Grant No. PHY-0098645. The numerical 
calculations have been performed on the IBM SP of the NIC, J\"ulich, Germany.

\bibliography{lit040504}

\end{document}